# Thermodynamic and kinetic fragility of Freon113: the most fragile plastic crystal

A. Vispa[1], M. Romanini[1], M. A. Ramos[2], L. C. Pardo[1], F. J. Bermejo[3], M. Hassaine[2], A.I. Krivchikov[4], J. W. Taylor[5], J. Ll. Tamarit[1,*]

[1] Grup de Caracterització de Materials, Departament de Física, ETSEIB, Universitat Politècnica de Catalunya, Diagonal 647, 08028 Barcelona, Catalonia, Spain

[2] Laboratorio de Bajas Temperaturas, Departamento de Física de la Materia Condensada, Condensed Matter Physics Center (IFIMAC) and Instituto Nicolás Cabrera, Universidad Autónoma de Madrid, Francisco Tomás y Valiente 7, 28049 Madrid, Spain

[3] Instituto de Estructura de la Materia, CSIC, Serrano 123, 28006 Madrid, Spain

[4] B. Verkin Institute for Low Temperature Physics and Engineering of NAS Ukraine, 47 Science Ave., 61103 Kharkov, Ukraine

[5] Rutherford Appleton Lab, ISIS Facility, Didcot OX11 0QX, Oxon, England and European Spallation Source, Data management and software centre, COBIS, Copenhagen 2200, Denmark

**Abstract**

We present a dynamic and thermodynamic study of the orientational glass former Freon113 (1,1,2-trichloro-1,2,2-trifluoroethane, $CCl_2F$–$CClF_2$) in order to analyze its kinetic and thermodynamic fragilities. Freon113 displays internal molecular degrees of freedom which promote a complex energy landscape. Experimental specific heat and its microscopic origin, the vibrational density of states from inelastic neutron scattering, together with the orientational dynamics obtained by means of dielectric spectroscopy have revealed the highest fragility value, both thermodynamic and kinetic, found for this orientational glass former. The excess in both Debye-reduced specific heat and density of states (*boson peak*) evidences the existence of glassy low-energy excitations. We demonstrate that early proposed correlations between the boson peak and the Debye specific heat value are elusive as revealed by the clear counterexample of the studied case.

[*] Corresponding author: josep.lluis.tamarit@upc.edu

When a structurally disordered system is rapidly cooled to avoid crystallization, some properties, as viscosity, show a dramatic increase down to the glass transition where the material reaches viscosity values comparable to those of a solid ($10^{12}$ Pa·s), i.e. relaxation times of $\approx 100$ s. Such a behavior contrasts with that typical for most liquids at high temperatures which usually exhibit a simple Arrhenius behavior of the relaxation time, $\tau = \tau_o \exp(E_a / k_B T)$, where the activation energy is temperature independent.

As the temperature is decreased the relaxation time shows a stronger increase, faster than that followed by the Arrhenius law and accompanied with an increase of some characteristic cooperativity relaxation length. The viscosity (or $\tau$) increase is generally characterized by recourse to the concept of the *kinetic* fragility [1,2], $m = \left[ \dfrac{\partial \log \tau}{\partial (T_g / T)} \right]_{T=T_g}$, which accounts for the deviation of the Arrhenius temperature dependence.

In terms of fragility index $m$, those materials for which relaxation times $\tau$ follow an Arrhenius law are known as "strong" glass formers, whereas "fragile" glass formers are those exhibiting super-Arrhenius behavior. For such cases, the temperature dependence of $\tau$ is given through the Vogel-Fulcher-Tammann (VFT) expression:

$$\tau = \tau_o \exp\left[ D \cdot T_o / (T - T_o) \right] \qquad (1)$$

where the temperature $T_o$ is associated with an ideal glass transition and even with the so-called Kauzmann temperature [3], and the fragility strength parameter $D$ is linked to the fragility parameter by $m = \dfrac{D \cdot T_o}{\ln 10 \cdot T_g} \cdot (1 - \dfrac{T_o}{T_g})^{-2}$ .

Typical strong glass formers ($m \approx 16$, or $D \geq 100$) are tetrahedral network liquids as $SiO_2$ or $GeO_2$. The highest values of fragility for organic materials (exception made of polymers) have been found in *cis/trans*-decahydro-naphthalene (decalin, $m=147$ [4]). Another group of materials exhibiting glass-like properties is that of crystals with positional order and orientational disorder [5]. Such plastic phases are formed from the liquid and can be supercooled giving rise to the so-called orientational glasses (OG) or "glassy crystals" [6-9]. They show typically low fragility, as cyclooctanol ($m=33$) [10,11], cycloheptanol ($m=22$) [12], ortho-carborane ($m=20$) [13], cyano-adamantane ($m=17$) [9,14], adamantanone ($m\approx16$) [9,15], ethanol ($m=48$) [9], or mixed molecular crystals $NPA_{0.7}NPG_{0.3}$ ($m=30$) [16-18]. The most fragile OG known to date are the Freon112 ($CCl_2F\text{-}CCl_2F$) with $m=68$ [19] and a co-crystal of succinonitrile (60%) and glutaronitrile (40%) with $m=62$ [20].

Several attempts to correlate the *kinetic* fragility associated with the relaxation time behavior as a function of temperature, either in liquids, polymers or plastic crystals, with their *thermodynamic* behavior [21-27] have been reported. A thermodynamic measure of fragility has also been defined [22] through the excess entropy $S_{exc}$ (the excess of liquid entropy over that of the crystal, that is generally taken as the configurational entropy $S_c$ [28] which appears in the Adams-Gibbs equation, $\tau \propto \exp[C/(TS_c)]$) scaled by the excess entropy at $T_g$, $S_{exc}(T_g)$. Such thermodynamic "Angell plots" ($T_g$-scaled Arrhenius plots) $S_{exc}(T_g)/S_{exc}(T)$ vs. $T_g/T$ exhibited a very similar behavior to that of classical Angell plots of the viscosity or $\tau$, log $\tau$ vs. $T_g/T$, for many glass-forming systems [25]. To quantify such correlation, a *kinetic* fragility $F_{1/2}$ is defined as $F_{1/2}=2T_g/T_{1/2}-1$, $T_{1/2}$ defined as the temperature half way between $10^{13}$ poise (or $10^2$ s for $\tau$) characteristic at $T_g$, and $10^{-4}$ poise (or $10^{-14}$ s), the high-temperature limiting value for liquids. Analogously, a thermodynamic fragility $F_{1/2}$ could be defined from the abovementioned $S_{exc}(T_g)/S_{exc}(T)$ normalized curves. Nevertheless, it was argued that for the *thermodynamic* fragility is preferable to use the $T_{3/4}$ line (at which $S_{exc}(T_g)/S_{exc}(T) = 3/4$) to avoid the need for extrapolations in the case of strong liquids. Hence $F_{3/4}=2T_g/T_{3/4}-1$. A good linear correlation between those kinetic and thermodynamic fragilities was claimed to be shown [25].

Compiled data for small organic molecule, polymeric, and inorganic glass-forming liquids by Huang and McKenna [24] revealed many deviations from the claimed correlation between kinetic and thermodynamic fragilities. By comparing the kinetic fragility index $m$ with the ratio of the liquid to the glass specific heats at $T_g$, $C_{p,liq}/C_{p,gl}$, as a measure of thermodynamic fragility, they confirmed the positive correlation between $m$ and $C_{p,liq}/C_{p,gl}$ for inorganic glass formers whereas opposite correlation was observed for polymeric glasses and $m$ was found to be nearly constant and independent of $C_{p,liq}/C_{p,gl}$ for small organic and hydrogen-bonding molecules. Nevertheless, this simple scaling of $C_p$ to assess thermodynamic fragility has been recently questioned [29]. Similarly, ref. [23] reports on the failure of the correlation for a set of molecular glass-formers by arguing that many-body molecular dynamics governing the kinetics are not embedded into a pure thermodynamic property as the entropy.

The correlation function $\phi(t)$ for the relaxation is described by the stretched exponential or Kohlrausch–Williams–Watts (KWW) function:

$$\phi(t) \approx \exp\left[-(t/\tau)^{\beta^{KWW}}\right] \qquad (2)$$

where the stretched exponent $\beta^{KWW}$ accounts for the departure of the exponential decay ($\beta^{KWW}=1$). The Fourier transform of the $\phi(t)$ function provides a good fit to the spectral shape of experimental relaxation data and it is closely related to the empirical Havriliak-Negami (HN) function [30]. The exponent $\beta^{KWW}$ is related to the frequency width of the asymmetric imaginary part of the susceptibility, accounting for the cooperative character of the relaxation.

Here we present a dynamic and thermodynamic study of the orientational glass former Freon113 (1,1,2-trichloro-1,2,2-trifluoroethane, $CCl_2F-CClF_2$) in order to analyze the kinetic and thermodynamic fragilities. Freon113 belongs to the series $C_2X_{6-n}Y_n$, with X, Y= H, Cl, F, Br, exhibiting plastic crystal phases [31-36]. These simple molecules display internal molecular degrees of freedom which promote the appearance of distinct conformers (trans and gauche) with low frequency internal modes that are able to couple with lattice motions, giving rise to a complex energy map [31,32]. Owing to these degrees of freedom, nucleation of the ordered phases is hindered and the high-temperature plastic phase is easily supercooled. The easiness to arrest these orientationally-disordered phases has indeed been linked to the existence of internal molecular degrees of freedom for a similar compound, Freon112 ($CCl_2F–CCl_2F$). Interestingly, for Freon112 the value for the fragility was reported to be the highest for a plastic crystal (*m*=68) [19]. In addition, our previous study on thermal conductivities of Freon112 and Freon113 compounds [35] shows that quasilocalized low-energy vibrational modes emerge at very low energy, lower than the values of the maximum of the *boson peak*, when compared to other OG. These low-energy modes in glassy systems promote an increase of the vibrational density of states $g(\omega)$ and, consequently, of the heat capacity excess ($C_{p,exc}$) over the Debye behavior $C_D \propto T^3$. Some authors [37] have indeed proposed a correlation between the $[C_{p,exc}]_{max}/C_D$ ratio, $[C_{p,exc}]_{max}$ being the maximum of the excess specific heat, and the fragility for glass-forming systems, the higher the fragility index, the smaller the ratio.

To quantify the thermodynamic fragility of Freon113, we have obtained the corresponding entropy curves for the OG/plastic crystal (crystal I) phases from specific-heat measurements. In Fig. 1, we depict our previously published data [35] around the glass-like transition between the OG and the plastic crystal, i.e. $T_g$=72 K.

The boson peak that dominates the low-frequency vibrational spectrum of glasses has been measured by both inelastic neutron scattering and specific heat. Debye-reduced vibrational density of states $g(\omega)/\omega^2$ and specific heat $C_p/T^3$ are plotted in Fig. 2 for both Freon112 and Freon113. The boson peak of Freon113 is higher and occurs at a higher energy (1.9 meV, 5 K) than that of Freon112 (1.5 meV, 4.5 K).

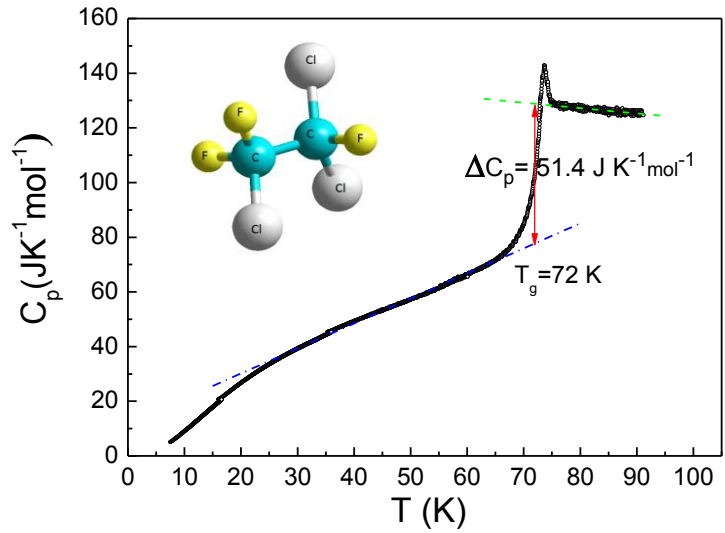

FIG. 1 (color online). Specific heat of Freon113 [34] around the glass-like transition between the orientational glass and the plastic crystal around $T_g$=72 K. Inset: scheme of the $CCl_2F$–$CClF_2$ molecule.

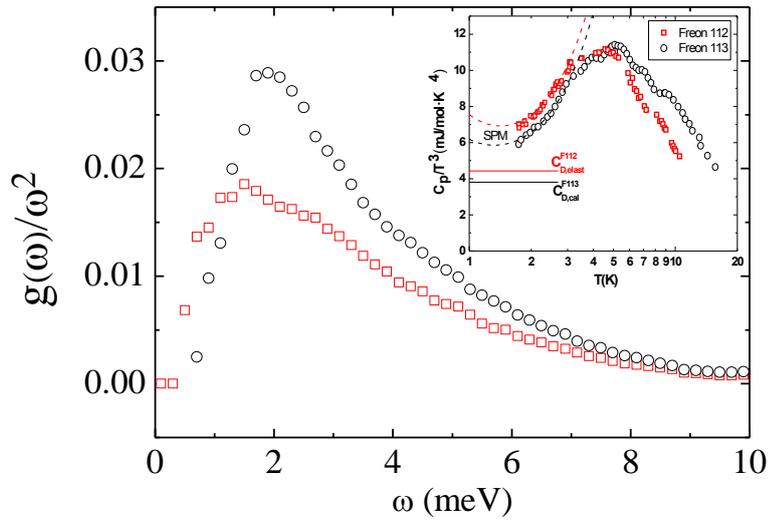

FIG. 2 (color online). Debye-reduced vibrational density of states divided by $\omega^2$ for Freon112 (red squares) and for Freon113 (black circles) determined at 5 K by means of inelastic neutron scattering at MARI spectrometer. Inset shows the molar heat capacity divided by $T^3$ for Freon112 and Freon113 [35], the calculated Debye values and the fits (dashed lines) corresponding to the soft potential model in a semi-log plot.

Kinetic properties of the plastic phase were determined through broadband dielectric spectroscopy. Fig. 3 shows the relaxation time of the different dynamic processes emerging in the imaginary part of the dielectric permittivity (inset in Fig. 3). The main non-Arrhenius α-relaxation process is accompanied by two slower processes associated with internal degrees of freedom. It is worth mentioning that according to ongoing molecular dynamic simulations, they are associated with intramolecular conformational changes which can couple to the orientational dynamics. The dielectric loss spectra were fitted according to a superposition of the Havriliak-Negami function (α-relaxation) with a $\beta^{KWW}$ exponent (calculated according to ref. [30]) ranged between 0.27 (at 70 K) and 0.62 (at 90 K), which strongly decreases with decreasing temperature, thus highlighting the increase of cooperativity due to strong orientational correlations between nearest neighbors. The temperature dependence of the α-relaxation time $\tau_\alpha$ is much more pronounced than a (simply activated) Arrhenius behavior, and it was modelled with the VFT eq. (1). $\tau_\alpha$ reaches $10^2$ s, the conventional relaxation time for glass transition, at 71±1 K, in close agreement with the thermodynamic glass transition temperature (72 ±1 K). The calculated kinetic fragility index provides a value of $m$=127, which is the highest so far reported for an OG, as can be seen in Fig. 4a in which $\tau_\alpha$ for several plastic crystals and some highly-fragile canonical glass formers are plotted as a function of $T_g/T$. As far as the *kinetic* fragility values $F_{1/2}$ and $F_{3/4}$ are concerned, 0.615 and 0.856 are respectively found.

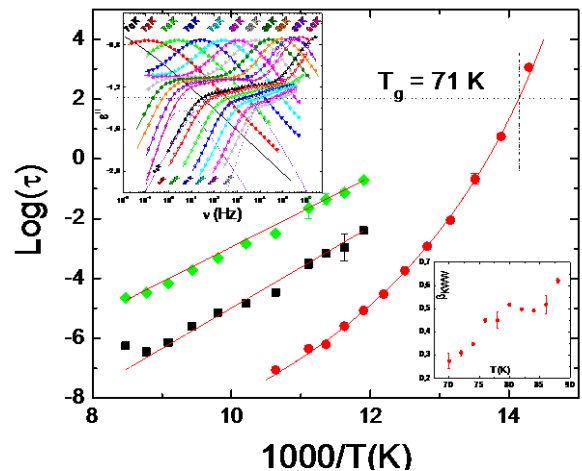

FIG. 3 (color online). Relaxation time as a function of reciprocal temperature for both α (red circles) and slower (green diamonds and black squares) relaxations. Top left inset shows obtained dielectric spectra for several temperatures ranging from 70 to 90 K. Solid lines are examples of the fits using Havriliak-Negami and Cole-Cole functions (dashed lines for the spectrum at 88 K) for the main α-relaxation and the slower processes, respectively. Bottom right inset collects the $\beta^{KWW}$ exponent as a function of temperature.

The entropy curve for the glassy crystal/plastic crystal phase below 125 K (inset of Fig. 4b) is obtained after numerical integration of the corresponding specific heat curve of Fig. 1, whereas the entropy data for the reference stable crystal II is taken from Kolesov et al. [38]. By subtracting the latter from the former one obtains the excess entropy

$S_{exc}(T)$ of the glassy phase. The so-obtained excess entropy for Freon113 is presented in Fig. 4b following the thermodynamic fragility plot introduced by Martinez and Angell [25]. Accordingly, Freon113 behaves as a very fragile glass-former thermodynamically too, with $T_g/T_{3/4}=0.988\pm0.004$ and hence $F_{3/4}=0.976\pm0.008$.

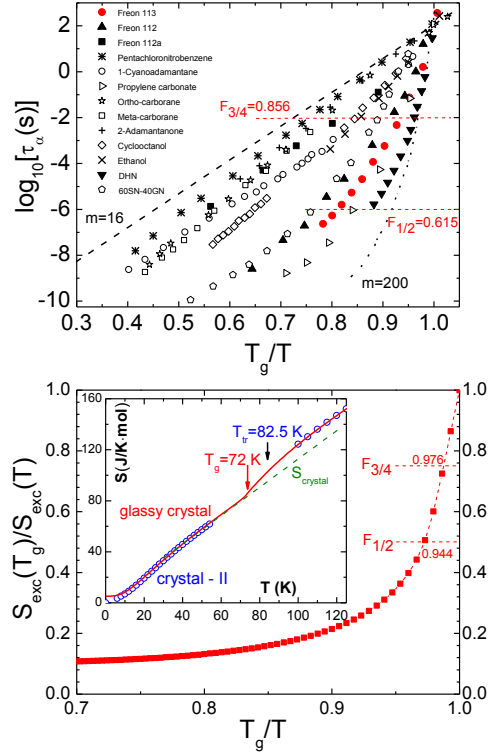

FIG. 4 (color online). (a) $\tau_\alpha$ as a function of $T_g/T$ (Angell plot). VFT parameters for Freon113 are: $T_0=58.9\pm0.6$ K, $\tau_o=(2.9\pm1.6)\cdot10^{-12}$, $D=(6.4\pm0.5)$. Fragility index is $m=127$. (b) Reciprocal of the excess entropy −normalized by the excess entropy at $T_g$− as a function of $T_g/T$. Inset: Entropy for the plastic/glassy crystal phase (red solid line), for the stable crystal II and the liquid (blue points, from [38]) as a function of temperature. The green dashed line indicates the extrapolation of the crystal II as a reference for higher temperatures. In both (a) and (b) figures, the dashed lines drawn at the ½ and ¾ marks are used to obtain either kinetic (a) or thermodynamic (b) fragilities $F_{1/2}$ and $F_{3/4}$, respectively.

The fragility of glass-forming liquids has been also correlated with their low-energy anomalous behavior in the glass state, as mentioned above. Sokolov et al. [37] found a negative correlation between the height of the boson peak in $C_p/T^3$ relative to the reference Debye level $c_{Debye}$, where $C_D = c_{Debye}T^3$, and the degree of fragility of the liquid, that is, the stronger the glass-forming liquid the higher its boson peak. Fig. 5 represents the data tabulated by Sokolov et al. [37], namely $C_{p,exc}/C_D = (C_p/C_D - 1)$ evaluated at the temperature where the maximum in $C_p/T^3$ occurs, as a function of the usual fragility index $m$ together with our current data for Freon113.

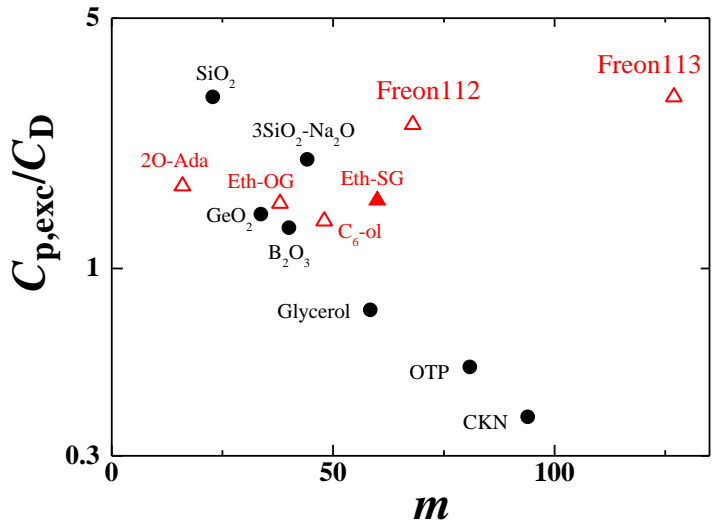

FIG. 5 (color online). Excess heat capacity $C_{p,exc} = C_p - C_D$, at the boson-peak maximum in $C_p/T^3$ scaled to $C_D$, as a function of the fragility index $m$ in a semi-log scale. Black circles correspond to canonical glasses, taken from the original figure from Ref. [37]. We have added some data (red triangles) from this (Freon113) and earlier works: adamantanone (2O-Ada) [39], cyclohexanol ($C_6$-ol) [40], Freon112 [31], and ethanol both as orientational (Eth-OG) and structural (Eth-SG) glass [41,42]. Solid symbols indicate structural glasses and open symbols indicate orientational glasses.

The clear excess of low-frequency excitations in the density of states (Fig. 2) for both Freon112 and Freon113, together with the evidence of their scattering with thermal acoustic phonons found through the thermal conductivity measurements [35], reveals the hybridization between intramolecular (torsional) degrees of freedom and acoustic phonon branches, an experimental fact that gives rise to the broad maximum in the reduced specific heat $C_p/T^3$ (inset in Fig.2). Within the framework of the multidimensional energy landscape [43,44], the potential energy function of the $N$-particle system represents the "topography" of its configurational space, the minima being the stable arrangements of the system. These minima, their number and their depth, depend on the specificities of the substance and, mainly, on the molecular structure and the inter- and intra-molecular interactions. An increase in the configurational entropy, due to the vibrational contributions or to the existence of other excitations, in excess over the crystal, as those coming from the different conformers and torsional degrees of freedom, should provide a faster decrease of the relaxation time (according to the Adam-Gibbs equation). As demonstrated by means of molecular dynamics simulations [45] fragility depends on changes in the vibrational properties of individual energy minima and how the number of minima is spread in energy. This would mean that the low-energy intramolecular excitations enhance such modes and consequently, increase the configurational entropy. The materials here studied are well representative of systems with a high value of the configurational entropy due to the intramolecular modes, clearly emerging at the dielectric susceptibility as slower processes than the main $\alpha$ relaxation (Fig. 3), as well as in the density of states (Fig. 2).

The Freon compounds here presented are representative cases of van der Waals molecular interactions but with strong short-range order [32] due to strong orientational correlations. These correlations are the result of the small energy difference between conformers but put apart by a large energy barrier, which strongly decreases with temperature, giving rise to strong coupling between these low-energy frequency modes and the lattice motions. A consequence of the strong orientational correlation is evidenced by the extremely low value of the $\beta^{KWW}$ exponent close to $T_g$ (inset in Fig. 3). Such a picture would indicate that, on the one hand, strong orientational correlation produces high kinetic fragility and, on the other, the low-energy excitations give rise to an extra contribution to the excess entropy. Moreover, the intramolecular force fields are known to contribute to the potential energy function and thus to impact on the energy landscape [44], since to the 3N dimensions (N being the number of particles) of the configuration space one must add the dimensions accounting for the asymmetry and non-rigidity of the involved molecular entities, that in the present case can become decisive to the profile of the energy landscape.

Moreover, the excess of low-energy frequency modes contributing to the excess heat capacity over the Debye value $C_D$ at the boson-peak maximum $C_{p,exc}$ (Fig.5) are not imperatively related to the acoustic modes of the ordered phase ($C_D$) and thus, correlation between $C_{p,exc}$ as a function of the fragility index $m$ can breakdown. Similar results were suggested by numerical simulations in a 2D glass-former system in which localized transverse vibrational modes associated with locally favored structures are responsible for the boson-peak [45], and by the study of metallic glasses [46] in which low-energy optical-like phonons or dispersive phonon branches can produce even larger peaks in scaled $C_p/T^3$ for the crystal than for its glassy counterpart.

Finally, we do not rule out that quantum effects for this low-temperature glass former can play an important role as claimed by Novikov and Sokolov [48] for glasses with $T_g$ near or below 60 K. For them, $T_g/T_m$ was predicted to be much smaller than the well-known classical value (2/3) due to the influence of zero-point vibrations on the glass transition. From literature data for molecular and hydrogen-bonded glass formers, the authors confirmed such a decrease of $T_g/T_m$ with decreasing $T_g$, and their proposed equation $T_g/T_m \approx A/[1+\theta_D/4T_g]$ fitted well the literature data with A≈0.8, though the lowest-temperature glasses tend to lie below the curve (see Fig. 2 of [48]). In the case of Freon 113 glassy crystal, $T_g$=72 K, $T_m$=238 K and $\theta_D$= 80 K [35], and hence $T_g/T_m$ = 0.30, in very good agreement with other glass formers of similar $T_g$ values. The activation energy, which is found to increase upon approaching $T_g$ for these low-temperature glasses, increases also in Freon 113 as usually found for normal glasses.

As a conclusion, we have demonstrated that the high values of kinetic ($m$=127) and thermodynamic ($F_{3/4}$ =0.976) fragilities of Freon113 (the highest values found in a plastic crystal to the best of our knowledge) are accompanied by a strong coupling between low-energy modes due to the high number of intramolecular degrees of freedom and acoustic phonon branches. Such modes contribute to the low temperature specific heat and to the density of states and, thus, to the excess (configurational) entropy, which makes the system thermodynamically and kinetically fragile. Finally we conclude that low-energy modes, which in an obvious way appear in the excess heat capacity over the Debye value at the boson-peak maximum, are in most cases uncorrelated with those of the ordered phase, making difficult a correlation between fragility and the excess specific heat, i.e., the boson peak.

This work has been partially supported by the Spanish MINECO through projects FIS2014-54734-P and FIS2014-54498-R and MAT2014-57866-REDT and "María de Maeztu" Programme for Units of Excellence in R&D (MDM-2014-0377). We also acknowledged the Generalitat de Catalunya under project 2014 SGR-581 and the Autonomous Community of Madrid through programme NANOFRONTMAG-CM (S2013/MIT-2850).